\documentclass[a4paper,11pt]{article}
\pdfoutput=1 

\usepackage{jheppub} 

\usepackage[OT1]{fontenc} 
\graphicspath{{Figures/}} 	     
\usepackage{booktabs} 
\usepackage{bbold} 
\usepackage{mathtools}
\usepackage{multirow}
\usepackage[version=4]{mhchem} 
\usepackage{physics}
\usepackage{pdflscape}
\usepackage{slashed}
\usepackage{xcolor}
\usepackage{caption}
\usepackage{ragged2e}
\usepackage{orcidlink}
\usepackage[normalem]{ulem}
\usepackage{makecell}

\newcommand{\bea} {\begin{eqnarray}}
\newcommand{\eea} {\end{eqnarray}}

\usepackage{graphicx}

\newcommand{\beq}{\begin{equation}}
\newcommand{\eeq}{\end{equation}}

\title{\Large 
R-parity violation and 8 TeV four-jet events at the LHC}
\vspace{4.5cm}

 \author[a,b]{Pedro Bittar \orcidlink{0000-0002-3684-5692},}
  \author[c]{Subhojit Roy \orcidlink{0000-0001-6434-5268},}
  \author[a,c,d,e,f]{and Carlos E.M. Wagner \orcidlink{0000-0001-6407-623X}}
\affiliation[a]{Perimeter Institute for Theoretical Physics, Waterloo, ON, N2L 2Y5, Canada}
\affiliation[b]{Department of Mathematical Physics,
Institute of Physics, 
University of São Paulo,
R. do Matão 1371, São Paulo, 
SP 05508-090, Brazil}
\affiliation[c]{HEP Division, Argonne National Laboratory, 9700 Cass Ave., Argonne, IL 60439, USA}
\affiliation[d]{Enrico Fermi Institute, Physics Department, University of Chicago, Chicago, IL 60637, USA}
\affiliation[e]{Kavli Institute for Cosmological Physics, University of Chicago, Chicago, IL 60637, USA}
\affiliation[f]{Leinweber Center for Theoretical Physics, University of Chicago, Chicago, IL 60637, USA} 
\emailAdd{bittar.hep@gmail.com}
    \emailAdd{sroy@anl.gov}
    \emailAdd{cwagner@uchicago.edu}
    \preprint{EFI-25-11} 
    
    \vspace{4.5cm}  

\abstract{
The CMS Collaboration at the Large Hadron Collider (LHC) has observed two four-jet events with a total invariant mass of about 8 TeV; within each event, the jets can be paired into two dijets with invariant masses of 2 TeV each. These are extremely rare events due to the large invariant mass, which implies a very small QCD background, as well as to the di-jet structure, which makes it prone to an interpretation in terms of a heavy resonance decaying into two lighter ones. We investigate the possible interpretation of these events in terms of supersymmetry with a single baryon-number and R-Parity violating term.
In this particular scenario, the lighter resonances are identified with the right-handed squarks of the first generation, while the heavy one is interpreted in terms of a down-squark of the second or third generation.  We discuss the constraints that shape this interpretation and outline a well-defined scenario for its realization. The resulting predictions can be scrutinized with forthcoming LHC data. }

\begin{document}
\maketitle
\flushbottom 
%
\section{Introduction}
Among the many possible extensions of the Standard Model (SM) at the TeV scale, supersymmetry has received considerable attention~\cite{Nilles:1983ge,Haber:1984rc,Martin:1997ns}. This is due mainly to the fact that supersymmetry allows for an extension of the perturbative and renormalizable description of the SM up to high energies, with a cancellation of the quadratic dependence of the Higgs
mass parameter on any possible heavy new physics scale, as well as the apparent unification of the three gauge couplings
in this scenario at a scale $M_{\rm GUT} \simeq 2 \times 10^{16}$~GeV, close to the Planck scale. 

Most of the studies of low energy supersymmetry have been done within the context of the conservation of R-Parity,
under which all SM particles (including the second Higgs doublet) are even, while all the supersymmetric partners
are odd. The presence of R-Parity suppresses proton decay and ensures that the lightest supersymmetry particle is
stable, being therefore a possible Dark Matter (DM) candidate. However, R-Parity is not necessary to remove the proton
instability. It is enough to suppress either the lepton-number or the baryon-number R-Parity operators. 
This is due to the fact that the proton is a fermion and the lightest baryon, and the only other known fermions lighter than the proton are leptons.  Therefore, any fermion number preserving proton decay channel must violate both baryon and lepton number. This conclusion can only be avoided in the presence of extra light fermions, like a light gravitino, something we will not consider in this work.  Moreover, although the inclusion of a light supersymmetric particle as DM is an attractive feature, there are many more DM candidates that may play this role~(see, for example, Ref.~\cite{Bertone:2016nfn}). 

R-Parity violation (RPV)~\cite{Dreiner:1997uz,Barbier:2004ez}, on the other hand, allows an easier interpretation of collider events that include no missing energy and many jets or leptons in the final state.
%
In this context, recently, the CMS collaboration at the LHC has detected two four-jet events with a very large invariant mass, of order 8 TeV.  These four jet events are built up by two di-jet events, each having an invariant mass of about 2~TeV~\cite{CMS:2022usq, CMS:2025hpa}. 
These events are very rare, and can be interpreted in terms of a heavy 8~TeV resonance decaying into two 2~TeV ones. 
Such four-jet signal can be interpreted in the context of diquark models as done in ~\cite{Dobrescu:2018psr,Dobrescu:2019nys,Dobrescu:2024mdl, Duminica:2025lte}. In particular, the model of Ref.~\cite{Dobrescu:2018psr} was later used in the CMS analysis \cite{CMS:2022usq, CMS:2025hpa}.
Due to the large invariant mass,  the heavy resonance must be produced by the collision of two valence quarks~\footnote{For a similar anomaly, at lower mass scales, see Ref.~\cite{Crivellin:2022nms, Dobrescu:2024mdl}}.
We propose to interpret it in terms of a single R-Parity violating coupling $\lambda_{11k}$, with
\begin{equation}
W_{RP} = \frac{\lambda^{''}_{ijk}}{2} \epsilon_{\alpha\beta\gamma} U^\alpha_iD^\beta_jD^\gamma_k ,
\label{eq:RPVcoup}
\end{equation}
where $W_{RP}$ is the R-Parity violating superpotential, the subscripts denote generations, the Greek indices are
associated with color, a summation over indices is understood, and $U,D$ are the up and down conjugate right-handed quark superfields, respectively.  The couplings $\lambda^{''}_{11k}=-\lambda^{''}_{1k1}$ are constrained by neutron ($n$) oscillations and dinucleon decays. We shall discuss these constraints later. Other flavor constraints, such as tree-level B meson decays and one-loop meson oscillations, require multiple RPV couplings to be sizable \cite{Giudice:2011ak} and are therefore not relevant for our scenario.

This article is organized as follows: in Sec.~\ref{collidersign} we introduce a minimal RPV SUSY setup, specify the masses and couplings that yield an $\sim 8~\mathrm{TeV}$ parent squark decaying to two $\sim 2~\mathrm{TeV}$ first-generation squarks, and show how this topology reproduces the observed four-jet excess. We present the associated collider phenomenology - production modes, widths, acceptances, and cross sections - and confront it with constraints coming from the di-jet and multi-jet searches at the LHC. Sec.~\ref{sec:noncollider} discusses non-collider constraints from $n\!-\!\bar n$ oscillations and dinucleon decays. Sec.~\ref{Conclusions} contains our conclusions and falsifiable predictions.

\section{Model and collider interpretation of the 8-TeV four-jet excess}
\label{collidersign}

\subsection{RPV Setup and assumptions}
The simplest way to resonantly produce a heavy state at the LHC is via valence-quark collisions, since their parton distribution functions (PDFs) are not strongly suppressed at large momentum fraction $x$. Up quarks are the most abundant, but there is no  RPV operator with two up quarks. In the $UDD$ scenario, the relevant coupling is $\lambda''_{11k}$ with $k=2,3$, so the $u d$ initial state can resonantly produce a right-handed anti-strange or anti-bottom squark $\tilde d_k^{\,*}$ of electric charge $+1/3$. We focus on purely hadronic final states (no leptons, negligible $E_T^{\text{miss}}$), concentrate on baryon-number violation through $\lambda''_{11k}$, and assume all other $UDD$ couplings are much smaller,
\begin{equation}
\lambda''_{mnl} \ll \lambda''_{11k}\,,\qquad (m,n,l)\neq (1,1,k)\,,
\end{equation}
see Ref.~\cite{Dercks:2017lfq} for a general collider overview. Throughout, we assume negligible left--right mixing in the sbottom sector. It is also worth recalling that third-generation squark masses in the $\sim(2\text{--}10)$~TeV range arise naturally in MSSM scenarios yielding a $125$~GeV Higgs~\cite{Ellis:1990nz,Haber:1990aw,Casas:1994us,Carena:1995bx,Carena:1995wu,Haber:1996fp,Heinemeyer:1998np,Espinosa:2000df,Carena:2000dp,Degrassi:2002fi,Degrassi:2012ry,Hahn:2013ria,Lee:2015uza,PardoVega:2015eno,Bahl:2017aev,Slavich:2020zjv}.

\subsection{Production, decay, and branching ratios}
To allow the heavy $\tilde d_k^{\,*}$ to decay into lighter squarks, we introduce the antisymmetric soft trilinear coupling $A_{ijk}=-A_{ikj}$ via
\begin{equation}
V \;=\; -\frac{A_{ijk}}{2}\,\epsilon_{\alpha\beta\gamma}\,\tilde u^\alpha_i\,\tilde d^\beta_j\,\tilde d^\gamma_k \;+\; \text{h.c.}\,,
\end{equation}
with the same flavor conventions as in the dimensionless RPV case, Eq.~(\ref{eq:RPVcoup}). The competing two-body widths are
\begin{align}
\Gamma\!\left(\tilde d_k^{\,*} \to \tilde u_i\,\tilde d_j\right) &= \frac{|A_{ijk}|^2}{8\pi\,m_{\tilde d_k}}\,
\sqrt{\left(1-x_i-x_j\right)^{\!2}-4x_i x_j}\,, \label{eq:widthA}\\[3pt]
\Gamma\!\left(\tilde d_k^{\,*} \to u\,d\right) &= \frac{|\lambda''_{11k}|^2}{8\pi}\,m_{\tilde d_k}\,, \label{eq:widthUD}
\end{align}
where $x_r \equiv m^2_{\tilde q_r}/m^2_{\tilde d_k}$. 
Since all squarks involved in this analysis are the superpartners of the right-handed quarks, we have omitted the right-handed subscript to simplify our notation, something we will also do in the rest of this article. We also assume that the up- and down-squarks involved in the $\tilde{d}_k$ decay are degenerate in mass, of order 2 TeV. 

It is important to note that once the coupling $\lambda^{''}_{11k}$ is fixed to obtain the proper production rate, the branching ratio for the decay of the $\tilde{d}_k$ squark to the lighter squarks $BR(\tilde{d}_k^* \to \tilde{u}_i\tilde{d}_j)$ may be controlled by the ratio  of the trilinear coupling $A_{ijk}$ to the $\tilde{d}_k$ mass, and assuming that all other supersymmetric particles are heavier than the squark $\tilde{d}_k$, it is naturally of order one. Diagrammatically, the resonant chain $u d \to \tilde d_k^{\,*} \to \tilde u_i\,\tilde d_j \to 4j$ is shown in the left panel of Fig.~\ref{fig:diagrams_LHC}.
The branching ratio that governs the four-jet topology is then
\begin{equation}
\mathrm{BR}\!\left(\tilde d_k^{\,*}\!\to\!\tilde u_i\,\tilde d_j\right)
=\frac{\Gamma(\tilde d_k^{\,*}\!\to\!\tilde u_i\,\tilde d_j)}{\Gamma(\tilde d_k^{\,*}\!\to\!\tilde u_i\,\tilde d_j)+\Gamma(\tilde d_k^{\,*}\!\to\!u\,d)}\,.
\label{eq:BR}
\end{equation}

For the benchmark choice $A_{11k}\!\sim\!5$~TeV and $\lambda''_{11k}\!\sim\!0.33$ with $m_{\tilde d_k}\!\simeq\!8$~TeV and $m_{\tilde u_i,\tilde d_j}\!\simeq\!2$~TeV, Eq.~\eqref{eq:BR} yields $\mathrm{BR}\!\approx\!0.75$. The corresponding total width is small, $\Gamma_{\tilde d_k}\!\sim\!142$~GeV ($\sim\!1.8\%$), consistent with a narrow-resonance treatment.

\subsection{Rates and collider constraints}
The $s$-channel production cross section of $\tilde{d}_k^*$ depends on the $u,d$ PDFs and scales as $|\lambda''_{11k}|^2$. Using {\tt MadGraph5\_aMC@NLO}~\cite{Alwall:2014hca} at LO with a $K$-factor of $1.3$ to emulate NLO QCD,

\begin{equation}
    \sigma\!\left(pp\to \tilde d_k^{\,*}\to 4j\right)
    \simeq 7.0\times 10^{-2}\ \text{fb}\;
    \left(\frac{\lambda''_{11k}}{0.33}\right)^{\!2}\!, 
    \label{eq:sigA}
\end{equation}

\noindent where scaling assumes $A_{11k}$ is adjusted to keep the decay branching ratio, Eq.~(\ref{eq:BR}), fixed. The subsequent $\tilde u_i,\tilde d_j\to jj$ decays yield the desired four-jet signal. For our masses, the off-shell contribution from the heavy squark is $\mathcal{O}(1/2)$ of the total rate. 
We set the acceptance based on the CMS analysis of narrow dijet resonances in Ref.~\cite{CMS:2025hpa}. CMS reports an acceptance$\times$efficiency of order \(20\%\) for a narrow-width diquark resonance with mass around 8~TeV after applying their dijet selection cuts. Re-implementing the same kinematic cuts on our simulated four-jet signal built from quarks, we verified that the fraction of events passing the selection is compatible with this acceptance. 
With a constant acceptance $A\simeq 0.2$ and $\mathrm{BR}\simeq0.75$, Eq.~\eqref{eq:sigA} implies about $2.5$ four-jet events for $\lambda''_{11k}\!=\!0.33$, i.e., roughly $1.44$ on-shell events plus $1.07$ off-shell event per experiment, to be compared with two on-shell and one off-shell candidate in CMS and one apparent off-shell candidate in ATLAS~\cite{ATLAS:2023ssk,CMS:2022usq,CMS:2025hpa}.
The relative size of the on-shell and off-shell contributions is obtained by comparing two {\tt MadGraph} samples: one for the full process and one where the intermediate squark is forced on-shell. The off-shell rate follows from subtracting the latter from the former. For the benchmark point, the two components are comparable, each contributing about \(50\%\) of the total.
A compact summary of the benchmark inputs, widths, cross sections, acceptance, and implied $4j$ yields is given in Table~\ref{tab:summary4j}. \\[3mm]

\begin{figure}[t]
    \centering
    \includegraphics[width=0.49\linewidth]{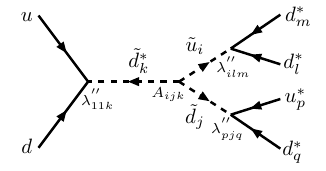}\hspace{3mm}
    \includegraphics[width=0.45\linewidth]{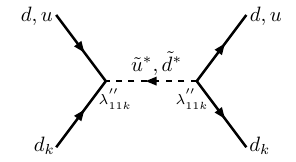}
    \caption{Representative diagrams for squark production and decay at the LHC. \textit{Left:} the 8~TeV $\tilde d_k^{\,*}$ decays via the soft $A_{ijk}$ into two $\sim2$~TeV first-generation squarks, yielding a fully hadronic four-jet signal with $(jj)(jj)$ substructure. \textit{Right:} leading constraint from resonant production of the light first-generation squarks, yielding dijet final states.}
    \label{fig:diagrams_LHC}
\end{figure}


\begin{table}[t]
  \centering
  \renewcommand{\arraystretch}{1.5}
  \begin{tabular}{|l  l  |  l l|}
    \hline
    \multicolumn{2}{|c|}{\textbf{Inputs}} & \multicolumn{2}{c|}{\textbf{Outputs}} \\
    \hline
    Parent mass & $m_{\tilde d_k}=8~\mathrm{TeV}$ 
      & Branching ratio & $\mathrm{BR}\simeq 0.75$ \\

    Daughter masses & $m_{\tilde u_i}, m_{\tilde d_j}\simeq 2~\mathrm{TeV}$
      & Total width & $\Gamma_{\tilde d_k}\simeq 142~\mathrm{GeV}$ \\[1.5mm]

    Couplings &\parbox[c]{3cm}{%
  $\lambda''_{11k}=0.33$\\
  $A_{11k}=5~\mathrm{TeV}$}
      &\parbox[c]{3cm}{\hspace{5mm}Signal \\ Cross Section} & $\sigma(pp\!\to\!4j)\simeq 7.0\times10^{-2}~\mathrm{fb}$ \\[1.5mm]

    Acceptance (4j) & $A\simeq 0.20$
      & $N_{4j}$ @ $139~\mathrm{fb}^{-1}$ & $\approx 2.5$ events \\

    & 
      & On/off-shell split & $\sim 1.4$ on-shell $+ 1.1$ off-shell \\
    \hline
  \end{tabular}
  \caption{Benchmark inputs and resulting $4j$ predictions for $(\tilde d_k^{\,*}\!\to\!\tilde u_i\,\tilde d_j\!\to\!4j)$.}
  \label{tab:summary4j}
\end{table}

\noindent\textit{Dijet searches.}—
Further constraints arise from resonant production of the light right-handed up- and down-type squarks via $u d \to \tilde d^{\,*},\tilde u^{\,*}\to q q$ (Fig.~\ref{fig:diagrams_LHC}, right). Our calculation agrees with Ref.~\cite{Pascual-Dias:2020hxo}. For $m_{\tilde q}\!\simeq\!2$~TeV we find

\begin{align}
\sigma(pp \to \tilde d^{\,*},\tilde u^{\,*}\to s\,u,\,s\,d)\,A &\simeq 152~\text{fb}\,
\left(\frac{\lambda''_{112}}{0.33}\right)^{\!2}\!, \label{eq:sigdijets2}\\
\sigma(pp \to \tilde d^{\,*},\tilde u^{\,*}\to b\,u,\,b\,d)\,A &\simeq \;68~\text{fb}\,
\left(\frac{\lambda''_{113}}{0.33}\right)^{\!2}\!, \label{eq:sigdijets3}
\end{align}

\noindent with an overall acceptance $A\simeq0.57$ as reported in the analysis of Ref.\,\cite{Pascual-Dias:2020hxo}.
Independently, we reproduced this value by generating a narrow dijet signal and applying the same kinematic cuts used in the CMS study of Ref.~\cite{CMS:2018mgb}.
The latest CMS (36~fb$^{-1}$) and ATLAS (139~fb$^{-1}$) dijet searches set upper limits of $\sim\!100$~fb and $\sim\!80$~fb, respectively, near 2~TeV~\cite{CMS:2018mgb,ATLAS:2019fgd}. Thus $\lambda''_{112}\!=\!0.25$ would already saturate the bound, implying a tighter inferred limit $\lambda''_{112}\!\lesssim\!0.26$, which in turn strains the four-jet rate demanded by Eq.~\eqref{eq:sigA}. In addition, much stronger constraints on $\lambda^{''}_{112}$ come from the bounds on dinucleon decays to kaons, as we discuss in Section \ref{sec:noncollider}. In contrast, for $k\!=\!3$, $\lambda''_{113}\!\sim\!0.33$ remains consistent with Eqs.~\eqref{eq:sigdijets3} and the data, allowing a simultaneous description of the four-jet signature. A corollary is the expectation of a \emph{resonant} dijet feature near $m_{jj}\!\sim\!2$~TeV at higher luminosity in the viable $k\!=\!3$ setup.
A numerical comparison with current limits and the resulting bounds on $\lambda''_{11k}$ is collected in Table~\ref{tab:dijetbounds}. \\[3mm]
\begin{table}[t]
  \centering
  \renewcommand{\arraystretch}{1.25}
  \caption{Dijet constraints near $m_{jj}\simeq 2~\mathrm{TeV}$ and implied bounds on $\lambda''_{11k}$ (acceptance $A\simeq 0.57$ included).}
  \label{tab:dijetbounds}
  \resizebox{1.0\linewidth}{!}{%
  \begin{tabular}{l c c c}
    \hline
    \textbf{Channel} & \textbf{Prediction at $\lambda''=0.33$} & \textbf{95\% CL limit} & \textbf{Implied bound on $\lambda''$} \\
    \hline
    $su,sd\!\to\!\tilde q_1^{\,*}\!\to su,sd$ ($k=2$) & $\sigma A \simeq 152~\mathrm{fb}$ & $80$--$100~\mathrm{fb}$~\cite{ATLAS:2019fgd,CMS:2018mgb} & $\lambda''_{112}\lesssim 0.26$ \\
    $bu,bd\!\to\!\tilde q_1^{\,*}\!\to bu,bd$ ($k=3$) & $\sigma A \simeq 68~\mathrm{fb}$  & $80$--$100~\mathrm{fb}$~\cite{ATLAS:2019fgd,CMS:2018mgb} & $\lambda''_{113}\lesssim 0.38$ \\
    \hline
  \end{tabular}}
\end{table}

\noindent\textit{Non-resonant $4j$.}—
Pair production of $m_{\tilde q}\!\simeq\!2$~TeV squarks also yields non-resonant four-jet final states. The cross section depends sensitively on the gluino mass. In the heavy-gluino limit (relevant $t$-channel propagators far off shell), the NLO rate is $\sim\!10^{-2}$~fb per right-handed flavor~\cite{Borschensky:2014cia}. Comparing with the current $13$~TeV bound on non-resonant high-mass $4j$ production~\cite{CMS:2022usq} for equal-mass squarks,
\begin{equation}
\sum_i \sigma(pp \to \tilde q_i\,\tilde q_i^*)\times A \;\lesssim\; 10^{-1}\ \text{fb}\,,
\end{equation}
and assuming all left-handed squarks are heavier than the heaviest right-handed state with at most four right-handed flavors near $2$~TeV, the predicted rate lies below present limits by factors of a few. This channel will therefore be testable at the HL-LHC.

Let's emphasize that although this is the most natural scenario for the explanation of the four jet events, one can find alternative scenarios within the same R-Parity violating framework. For instance,  to evade the strong dijet constraints, one can assume the first generation squarks to be heavy and the second generation strange and charm squarks to be the 2~TeV resonances involved in the heavy squark decays. In such a case, one can still use the dimensionless $\lambda^{''}_{113}$ coupling to produce the heavy 8 TeV~sbottom, but the dimensionful coupling allowing for the sbottom decay into lighter squarks should be $A_{223}$ instead of $A_{113}$. This would alleviate the bounds on $\lambda^{''}_{113}$ and allow for slightly larger sbottom masses, if needed to fit the data. 
The decay of the second generation squarks could be associated with a very small $\lambda^{''}_{ij2}$ coupling, which should be sufficiently large to ensure prompt sbottom decays.  The four jet signatures would
be similar to in the previous scenario, although the presence of bottom-quarks in the final state is not ensured. Assuming that $\lambda^{''}_{113}$ is the only relevant dimensionless R-Parity violating coupling, the resonant 2~TeV second generation squark production cross section will be highly suppressed. In particular, the couplings $\lambda^{''}_{212}$, $\lambda^{''}_{132}$,
and most importantly $\lambda^{''}_{112}$, which would allow the resonant light squark production via the collision of at least one valence quark, should be small, acquiring values of order 0.1 or smaller for this to happen.

\subsection{Four-jet mass distribution and 
discussion}
\begin{figure}[t]
    \centering
    \includegraphics[width=0.7\linewidth]{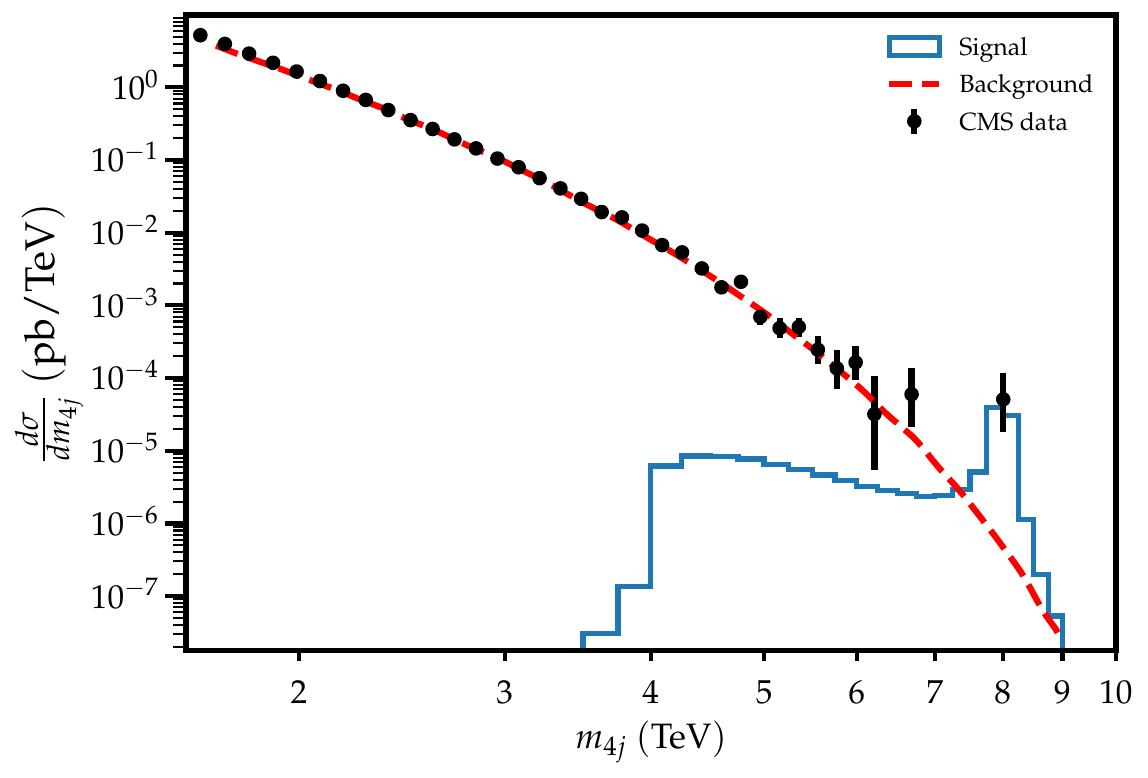}
    \caption{Four-jet invariant-mass distribution $m_{4j}$ comparing our signal prediction 
    (blue histogram) with the CMS measurements (black markers) and the SM 
    background estimate obtained from interpolation (red dashed)~\cite{CMS:2022usq,CMS:2025hpa}. 
    This distribution corresponds to the CMS inclusive analysis, where all 
    $\alpha = m_{2j}/m_{4j}$ bins have been combined.  At $m_{4j}=8~\text{TeV}$, the CMS data bin contains two overlapping data points.
    }
    \label{fig:4j_distro}
\end{figure}
We overlay our simulated signal on the CMS four-jet invariant-mass spectrum, 
$m_{4j}$, adopting the same binning and kinematic selections as the CMS 
four-jet searches~\cite{CMS:2022usq,CMS:2025hpa}. 
In Fig.~\ref{fig:4j_distro}, the blue histogram shows our signal prediction; 
black markers denote the CMS data; and the red dashed line is the SM background 
estimate obtained from the interpolation procedure used by CMS. 
At $m_{4j}=8~\mathrm{TeV}$, the CMS bin contains two overlapping data points. 

A central ingredient of the CMS strategy is the kinematic ratio
$\alpha \equiv m_{2j}/m_{4j}$,
constructed from the average dijet mass $m_{2j}$ and the four-jet mass $m_{4j}$. 
Binning the data in $\alpha$ yields a smoothly and steeply falling one-dimensional $m_{4j}$ spectra in each bin (mitigating sculpting from phase space), which 
suppresses the multijet background while leaving a genuine resonance localized 
in a small set of $\alpha$ bins (See Fig.~7 of Ref.~\cite{CMS:2022usq} and Fig.~5 of Ref.~\cite{CMS:2025hpa}). This directly improves the signal-to-background 
ratio in the bins where a resonance would appear.

In the CMS inclusive spectrum, two events with $m_{4j}\simeq 8~\mathrm{TeV}$ 
and $m_{2j} \simeq 2~\mathrm{TeV}$ drive a local (global) 
significance of $3.9~(1.6)\sigma$ for a narrow resonance near 
$M_Y\simeq 8.6~\mathrm{TeV}$ and $M_X\simeq 2.1~\mathrm{TeV}$~\cite{CMS:2022usq}. 
A similar analysis was done by ATLAS 
reported no events. 
In their analysis, they observe one event 
with $m_{4j} \sim 6.6$~TeV and $m_{2j} \sim 2.2$~TeV with no statistically 
significant excess overall.
Given uncertainties and low statistics, the ATLAS observation neither confirms nor excludes the new physics interpretation of the observed excess around $m_{4j} \sim 8$~TeV by the CMS collaboration\cite{CMS:2022usq, CMS:2025hpa}.

The observed excess is concentrated near $\alpha \sim 0.25$, consistent with a topology 
peaking at $m_{4j}\sim 8~\mathrm{TeV}$ and $m_{2j}\sim 2~\mathrm{TeV}$. 
Consequently, $\alpha$ selections that target $\alpha\approx 0.25$ preferentially reduce the QCD multijet background while retaining the signal, as explicitly seen in the CMS bin-by-bin presentation where the separation from background is enhanced compared to the inclusive view~\cite{CMS:2022usq}. 
In this work, for simplicity, we use the inclusive data combining all $\alpha$ bins (See Fig.~7 of Ref.~\cite{CMS:2025hpa}), $m_{4j}$ distribution in Fig.~\ref{fig:4j_distro} to 
confront our benchmark. 
In fact, ATLAS and CMS observed events with $m_{4j}$ below 8~TeV could be interpreted as the off-shell events in our scenario, as we expect one event, as pointed out in Tab.~\ref{tab:summary4j}.
Nevertheless, given the 
localization of our signal in $\alpha$, a CMS-style bin-by-bin analysis would 
be expected to yield a higher significance than the inclusive treatment, 
mirroring the improvement seen by CMS upon $\alpha$ binning. 
With $\lambda''_{11k}\!\sim\!0.33$ and $A_{11k}\!\sim\!5~\mathrm{TeV}$, our 
benchmark predicts $\mathcal{O}(1\text{--}2)$ events in the 
$m_{4j}\!\approx\!8~\mathrm{TeV}$ bin-compatible with the two CMS candidates 
featuring $m_{2j}\!\approx\!2~\mathrm{TeV}$ in each-while the 
interpolation-based SM background in that bin is very small. 
Additional luminosity will therefore provide a sharp test of this scenario. 
Moreover, since our framework can also generate excesses in different $\alpha$ regions, 
it would remain a viable explanation if future data reveal localized deviations outside 
the currently observed $\alpha \sim 0.25$ bin. In the $k=3$ case, two of the four jets 
originate from $\tilde b$ decays and should be $b$-taggable, offering an orthogonal handle.

\section{$n-\bar{n}$ oscillations and dinucleon decay bounds}
\label{sec:noncollider}
In the case of baryon- and R-parity violating couplings, one could violate baryon number by two units, leading to a contribution of $n-\bar{n}$ oscillations~\cite{Calibbi:2016ukt,Leontaris:2018blt}. This contribution may only be obtained in the presence of gluinos, and the amplitude depends on the gluino Majorana mass that allows the proper fermion anti-fermion conversion of the down quark not involved in the production of the $\tilde{d}_k^*$.  In addition, a mixing between the bottom squark, in the case $k=3$, and the down squark is required, to ensure the existence of the gluino vertex, implying that the amplitude
depends quadratically on this mixing.  

The bound can be written also in terms of the matrix element of the operator $(u_R d_R d_R)^2$ arising after integration of the heavy squarks and gluinos, between $n$ and $\bar{n}$ states~\footnote{ ($u_R d_R d_R$) refers to the color contracted operator)},
\begin{equation}
O_{n\bar{n}}=\langle  \bar{n} | (u_R d_R d_R)^2 | n \rangle  \sim {\cal{O}}(\Lambda_{\rm QCD}^6).
\end{equation}
The precise value of this operator is quite uncertain, and different estimates differ by an order of magnitude.  To get an idea of the bound on the R-Parity violating coupling, we quote the values presented in Ref.~\cite{Calibbi:2016ukt}, adapted to our case, 
\begin{equation}
\tau_{n\bar{n}} \sim  4.7 \times 10^8 s  \frac{m_{\tilde{g}}}{15 {\rm TeV}} \left(\frac{m_{\tilde{b}}}{8 \ {\rm TeV}}\right)^2 \left( \frac{m_{\tilde{d}}}{2 \ {\rm TeV}}\right)^2 \left(\frac{0.33}{\lambda^{''}_{11k}} \right)^2 \frac{(250 {\rm MeV})^6}{O_{n\bar{n}}}\left(\frac{6.3 \times 10^{-6}}{(\delta^d_{RR})_{k1}} \right)^2
\label{eq:nnbar}
\end{equation}
This implies that to get a neutron oscillation lifetime larger than $4.7\times 10^8$ s, as required by experiment \cite{ParticleDataGroup:2024cfk,Super-Kamiokande:2020bov}, one has to suppress either the coupling $\lambda^{''}_{11k}$ or the mixing $(\delta^d_{RR})_{1k}$. To satisfy the dinucleon decay bounds, $\lambda^{''}_{112}$ has to be suppressed as we discuss below. Then, to realize our collider scenario we keep the coupling $\lambda^{''}_{113} \sim 0.33$ and the 
right-handed sbottom-sdown mixing $(\delta^d_{RR})_{13}$ should be smaller than about $6.3\times 10^{-6} $.  There are, of course large hadronic
uncertainties associated with this computation, but this suppression would demand either no flavor mixing in the
quark right-handed sector, or a symmetry protecting the mixing of the first two generations with the third one in the right-handed down quark sector.
Otherwise, one should invoke an accidental cancellation of the mixing operator, which, however due to the magnitude
of  $(\delta^d_{RR})_{31} \simeq{\rm few} \ 10^{-6}$ does not look very likely. Independently of its origin, this mixing
cancellation is required in order for this scenario to survive experimental bounds. 

Additional bounds on the R-Parity violating couplings may be obtained by considering dinucleon decay into either pairs of pions or Kaons. The decay into Kaons is particularly efficient in constraining the coupling $\lambda^{''}_{112}$, which is already constrained by collider constraints. 
The dinucleon decay lifetime should be larger than $4 \times 10^{32}$ years \cite{Takhistov:2016eqm,Litos:2010zra}, leading to the bound,
\begin{equation}
\tau_{nn \to \pi\pi} \simeq  4 \times 10^{32} {\rm yrs}  \left(\frac{m_{\tilde{g}}}{12 {\rm TeV}}\right)^2 \left(\frac{m_{\tilde{b}}}{8 {\rm TeV}}\right)^4 \left( \frac{m_{\tilde{d}}}{2 {\rm TeV}}\right)^4 \left(\frac{3\times 10^{-5}}{\lambda^{''}_{112}} \right)^4 \frac{(150 {\rm MeV})^5}{O_{nn\pi\pi}} ,
\label{eq:dinucleon112}
\end{equation}
where we have assumed a nuclear matter density $\rho_N \sim 0.25 \ {\rm fm}^{-3}$ and
\begin{equation}
O_{nn\pi\pi}= \langle nn | (u_R d_R d_R)^2 | \pi\pi \rangle .
\end{equation}

\noindent Imposing the dinucleon decay limit requires $\lambda^{''}_{112} < 3\times 10^{-5}$.  
With this condition, the $n-\overline{n}$ oscillation bound from Eq.~\eqref{eq:nnbar} can be reinterpreted, which translates into a weaker constraint on the first-generation mixing: $(\delta^d_{RR})_{12}<0.07$.

The coupling $\lambda^{''}_{113}$,  is also constrained by the decay into pions that depend on parameters similar to the neutron oscillation case. Differently from \eqref{eq:dinucleon112}, in the case for $\lambda^{''}_{113}$, the decay lifetime also depends on the mixing of the first and third generation squarks. Applying the same bound, one obtains, 
approximately
\begin{equation}
\tau_{nn \to \pi\pi} \simeq  4 \times 10^{32} {\rm yrs}  \left(\frac{m_{\tilde{g}}}{12 {\rm TeV}}\right)^2 \left(\frac{m_{\tilde{b}}}{8 {\rm TeV}}\right)^4 \left( \frac{m_{\tilde{d}}}{2 {\rm TeV}}\right)^4 \left(\frac{0.33}{\lambda^{''}_{113}} \right)^4 \frac{(150 {\rm MeV})^5}{O_{nn\pi\pi}}\left(\frac{ 9 \times10^{-5}}{(\delta^d_{RR})_{31}} \right)^4 \, .
\end{equation}
It is clear from here that this decay provides less stringent limits on the mixing $(\delta^d_{RR})_{31}$ than the neutron oscillation lifetime. Similar constraints appear from the decay into Kaons, although the bound on the lifetime is a little bit smaller and the constraint is on the mixing of the right handed sbottom with the second generation right handed squark, which is therefore  bounded to be $(\delta^d_{RR})_{32} \lesssim 0.9\times 10^{-5}$. 

Altogether, for the collider scenario to work, the right-handed mixings with the third generation must be strongly suppressed, $(\delta^d_{RR})_{13}\lesssim 7\times 10^{-6}$ and $(\delta^d_{RR})_{23}\lesssim 0.9\times 10^{-5}$. The dinucleon decay requires $\lambda''_{112}\lesssim 3\times 10^{-5}$, which allows the first–second generation mixing $(\delta^d_{RR})_{12}$ to be relatively large, of order 0.07. This pattern points to a flavor structure that distinguishes the third generation from the first two, effectively making the dangerous low-energy processes under control. A natural possibility is to impose a $U(2)\times U(2)$ flavor symmetry in the right-handed sector. In this setup, the third generation is neutral under the symmetry while the first two generations are charged  and affected by a universal soft breaking mass parameter ensuring the approximate equality of the up and down squark masses of the first two generations, and preventing dangerous mixings. Small effects from hypercharge and Yukawa couplings break the symmetry slightly, but as long as only $\lambda''_{113}$ is relevant, no significant mixing is generated, keeping the scenario safe from low-energy~bounds.

\section{Conclusions}
\label{Conclusions}
The CMS collaboration has reported two four jet events with intriguing similar characteristics and that suggest the production of a heavy resonance, of mass about 8~TeV, decaying into lighter ones of order 2~TeV.
This explanation demands a baryon-number violating coupling of order 0.33, with a right-handed sbottom of mass about 8~TeV, while the first and/or second generation squarks must have masses of about 2~TeV.

The requirement of being in agreement with neutron oscillation and dinucleon decay constraints puts severe bounds on the possible mixing of the first and second right-handed squarks with the third generation right-handed squark. Barring possible accidental cancellations, this could be achieved by demanding, for example, an extended global symmetry under which only the right-handed quarks/squarks of the first two generations are charged. At the same time, a universal soft supersymmetry breaking parameter affecting this sector and ensuring the equality of the up and down squark masses should be present. In either way, this scenario cannot be excluded by these constraints.  The validity of the explanation presented in this work would require a few more 8~TeV~four jet events at higher LHC luminosities, in the near future. Moreover, in the main benchmark presented in this work two of the four jets should be tagged as bottom quarks and this may provide a further test of this scenario. 

\acknowledgments
The authors thank Bogdan Dobrescu for pointing out the CMS reported heavy resonance excess during a Fermilab Journal Club discussion.
PB is supported by Perimeter Institute for Theoretical Physics and was supported by Fundação de Amparo à Pesquisa do Estado de São Paulo (FAPESP) grant number 2025/03848-8, project 2021/11489-7, during the development of this project. Research at Perimeter Institute is supported by the Government of Canada through the Department of Innovation, Science and Economic Development Canada and by the Province of Ontario through the Ministry of Research, Innovation and Science.
SR is supported by the U.S.~Department of Energy under contracts No.\ DEAC02-06CH11357 at the Argonne National Laboratory.  He thanks the University of Chicago, Fermilab and the Perimeter Institute for Theoretical Physics, where a significant part of this work was carried out.
SR also thanks CW for hosting him at the Perimeter Institute, where the project was originally initiated.
The work of C.W. at the University of Chicago has been supported by the DOE grant DE-SC0013642. The work of Perimeter Institute has been supported by a Distinguished Visiting Research Chair position.

\bibliography{RPV}
\bibliographystyle{JHEP}

\end{document}